\documentclass[rnote]{aa} 

\usepackage{hyperref}
\usepackage{graphicx}
\usepackage{txfonts}
\begin{document}

   \title{A catalogue of photometric redshifts for the SDSS-DR9 galaxies}

   \author{M. Brescia \inst{1} \and S. Cavuoti \inst{1} \and G. Longo \inst{2} \and V. De Stefano \inst{2}
          }
\institute{INAF Astronomical Observatory of Capodimonte, via Moiariello 16, I-80131 Napoli, Italy\\
\email{brescia@oacn.inaf.it}
\and
University Federico II, Department of Physics, via Cinthia, I-80126 Napoli, Italy
\\
             }

   \date{Received May 2014; accepted July, 2014}


  \abstract
   {Accurate photometric redshifts for large samples of galaxies are among the main products of
modern multiband digital surveys. Over the last decade, the Sloan Digital Sky Survey (SDSS) has become
a sort of benchmark against which to test the various methods.}
   {We present an application of  a new method to the estimation of photometric redshifts for the galaxies in the SDSS
Data Release 9 (SDSS-DR9). Photometric redshifts for more than $143$ million galaxies were produced and made available at the URL: \\
\noindent \url{http://dame.dsf.unina.it/catalog/DR9PHOTOZ/}.}
   {The MLPQNA (Multi Layer Perceptron with Quasi Newton Algorithm) model provided within the framework of the DAMEWARE (DAta Mining and Exploration Web Application  REsource) is
an interpolative method derived from machine learning models. }
   {The obtained redshifts have an overall uncertainty of $\sigma=0.023$ with a very small average bias of $\sim 3\times 10^{-5}$, and a fraction of catastrophic outliers ($\left| \Delta z \right| > 2\sigma$) of $\sim 5 \%$.
   This result is slightly better than what was already available in the literature, also in terms of the smaller fraction of catastrophic outliers.}
   {}


   \maketitle
%
\section{Introduction}
In the last few years, photometric redshifts (photo-z) for large samples of normal or active galaxies have become crucial for a variety of cosmological applications (\citealt{scranton2005,myers2006,hennawi2006,giannantonio2008}) and many different methods for their evaluation have been presented and extensively discussed in the literature (cf. \citealt{hildebrandt2010}).
The problem of deriving accurate photometric redshifts has become even more cogent due to the
huge amount of data produced by most ongoing and planned photometric surveys (cf. PANNSTARS: \citealt{kaiser2004}, KIDS\footnote{\url{http://www.astro-wise.org/projects/KIDS/}}, EUCLID: \citealt{laureijs2011}) aimed at explaining weak lensing to prove the dark components of the universe.

Without entering into details which can be found elsewhere, it is worth reminding that, broadly speaking,
all photo-z methods are based on the interpolation of some a priori knowledge represented by sets of
templates, and differ only in one or both of the following aspects:
\textit{(i)} the way in which the a priori Knowledge Base (KB) is constructed (higher accuracy spectroscopic redshifts or, rather, empirically or theoretically derived Spectral Energy Distributions or SEDs), and \textit{(ii)} the interpolation/fitting algorithm employed.

In all methods, the main source of uncertainty is in the fact that the function mapping the color space into the spectroscopic redshift space is just an oversimplified approximation of the complex and otherwise unknown relation existing between colors and the redshift (as an example, see \citealt{csabai03}).
Among the various interpolative methods, we shall just quote a few: \textit{i)} polynomial fitting \citep{connolly95};
\textit{ii)} nearest neighbors \citep{csabai03};
\textit{iii)} neural networks (\citealt{dabrusco07,yeche10} and references therein);
\textit{iv)} support vector machines \citep{wadadekar05};
\textit{v)} regression trees \citep{carliles2010};
\textit{vi)} gaussian processes \citep{way06,bonfield10}, and
\textit{vii)} diffusion maps \citep{freeman09}.

In this paper we focus on the application to the galaxies contained in the SDSS Data Release $9$ (DR9, \citealt{dr9}), of the MLPQNA (Multi Layer Perceptron with Quasi Newton Algorithm) method already described in detail elsewhere \citep{brescia1, brescia2}, hence we refer the interested readers to these papers for all the mathematical and technical details.
We wish to notice that in the framework of the PHAT1 contest \citep{hildebrandt2010}, which blindly compared most existing methods for photo-z evaluation, the MLPQNA method proved to be among the two best empirical methods to date \citep{cavuoti0}. This in spite of the very limited base of knowledge available for the contest ($\sim500$ objects only).

MLPQNA is just one among the many data mining methods publicly available under the DAta Mining \& Exploration Web Application REsource infrastructure (DAMEWARE; \citealt{brescia3}).

In the next section (Sect.~\ref{thedata}) we describe the data set used as knowledge base, while in Sect.~\ref{experiments}
we describe the experiments and discuss their outcome. In Sect.~\ref{catalog} we describe the resulting catalogue of photometric redshifts and a short summary of the work is given in Sect.~\ref{conclusions}.

\section{The Data}\label{thedata}
The Sloan Digital Sky Survey (SDSS, \citealt{sdss}), is the forerunner of modern wide-field surveys. It combines multi-band photometry and fiber-based spectroscopy, thus providing both photometric data for a very large number of objects and spectroscopic information for a smaller but still significant subsample of the same population. Hence it provides all information needed to constrain the fit of an interpolating function mapping the photometric features into the spectroscopic redshift space.
This is the main reason why most, if not all, photometric redshifts methods have been tested on the various data releases of the SDSS which, over the years, has become a sort of \textit{benchmark} data set against which to test old and new methods.

To form our Knowledge Base (KB) we extracted from the spectroscopic subsample of the SDSS-DR9 all objects with specClass \textit{galaxy} together with their photometry, in particular we used the ($psfMag$) magnitudes and the related colors, rejecting all objects with missing or non detected information in any of the SDSS photometric bands.

\begin{figure*}
   \centering
   \includegraphics[width=14cm]{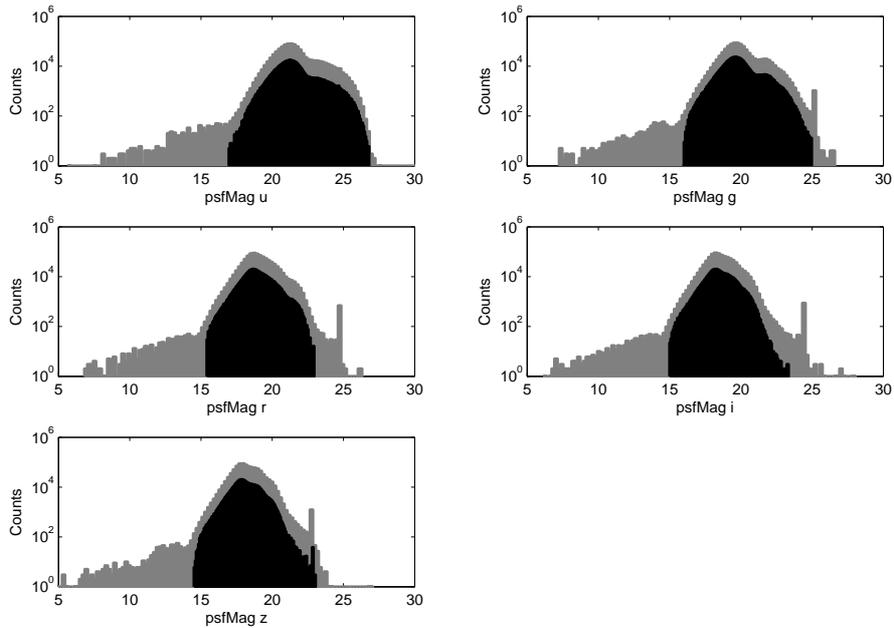}
   \caption{Distributions of the five $psfMag$ type magnitudes. For each magnitude the light gray area is referred to the whole spectroscopic sample, while the black one represents the objects sampled in on knowledge base.}\label{magdistrib}
    \end{figure*}

 The cuts in the magnitude were obtained by considering the limits within which the photometric parameter space sampled by the spectroscopic objects is significantly covered. Within these limits, the neural algorithm during the training phase is exposed in every regions of the cleaned parameter space to a number of examples sufficiently large to allow learning. Obviously the less populated will be the region of the parameter space, the less accurate is expected to be the accuracy of the final result. An additional implication is that the less populated is a region of the parameter space, the less likely is the capability to correctly learn the rule for peculiar or rare objects. The resulting psf magnitude limits are listed in Table~\ref{tabmaglim}. While Fig.~\ref{magdistrib} shows the psf magnitude distributions in the knowledge base. As also described in \citealt{oyaizu2008} we trained our model on the spectroscopic sample up to the magnitude limit of $psfMag\_r < 23.0$. By considering a photometric limit of $r<22.2$, the resulting fainter limit in the training set covers the complete photometric region of interest without introducing boundary effects for photometric redshifts of galaxies having magnitudes near the $psfMag\_r$ limit.
All this taken into account the complete spectroscopic KB consisted of $497,339$ objects.

\begin{table*}
\begin{center}
\begin{tabular}{lcc}
Band & lower limit & upper limit\\
u & 16.97 & 26.79\\
g & 15.91 & 24.97\\
r & 15.41 & 22.94\\
i & 15.00 & 23.34\\
z & 14.17 & 23.04\\
\end{tabular}
\end{center}
\caption{The $psfMag$ type magnitude limits derived in each band during the knowledge base definition.}\label{tabmaglim}
\end{table*}

\section{Experiments and discussion}\label{experiments}
In machine learning supervised methods it is common practice to use the available KB to build
at least three disjoint subsets for every experiment: one (training set) for training purposes,
i.e. to train the method in order to acquire the hidden correlation among the input features which
is needed to perform the regression; the second one (validation set) to check the training, in
particular against a loss of generalization capabilities (a phenomenon also known as overfitting);
and the third one (test set) to evaluate the overall performances of the model \citep{brescia2}.

In this work, the validation was performed during training, by applying the standard leave-one-out k-fold cross validation
mechanism \citep{geisser1975}. We would like to stress that none of the objects included in the training (and validation) sample was included
in the test sample and only the test data were used to generate the statistics.
In other words, the test was blind, i.e. based only on objects never submitted to the network.

\begin{figure*}
   \centering
   \includegraphics[width=14cm]{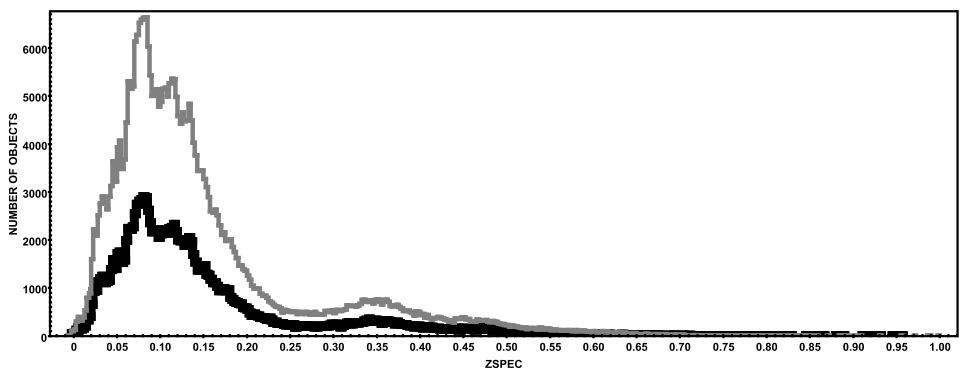}
   \caption{spectroscopic redshift distribution of objects included in the training set (black line) and test set
	(gray line).}\label{histospec}
\end{figure*}

We decided to populate the training and the test set with respectively $30$\% and $70$\%  of the objects in the KB, namely with $149,997$ and $347,342$ objects, respectively. This decision, which might seem a little anomalous since it is common practice for machine learning methods to operate with data sets of reversed proportion, was dictated by the large number of examples present in the knowledge base and by the specificity of the MLPQNA method which can \textit{overfit} (with a loss of generalization capability) the data when exposed to a very large number of examples.
The histogram in Fig.\ref{histospec} shows the distribution of the objects in the KB as a a function of the zspec in both training and test sets.

In order to ensure that the KB provided a proper coverage of the Parameter Space, the data were split into the two data sets by
random extraction. In other words, by randomly shuffling and splitting the original dataset, we replicated several times the extraction sequence, and evaluated the average of their output.
This mechanism prevents possible biases induced by fluctuations in the coverage of the parameter space, namely small differences
in the redshift distribution of training and test samples used in the experiments.

Once the data sets were produced, we checked which types of flux combinations were more effective, in terms of magnitudes or
related colors and therefore we performed and compared two experiments with two different sets of features: \textit{(i) MAG},
by using the five SDSS $psfMag$ magnitudes and \textit{(ii) MIXED}, by replacing the magnitudes with the derived colors and
leaving only the $r$ $psfMag$ as pivot magnitude. The best combination turned out to be the \textit{MIXED} type.
From the data mining point of view this is rather surprising since the amount of information should not change by applying
linear combinations between features. From the physical point of view, however, the better performances of the MIXED
experiment can be easily understood by noticing that even though colors are derived as a subtraction of magnitudes,
the content of information is quite different, since an ordering relationship is implicitly assumed, thus increasing
the amount of information in the final output (i.e., flux gradients instead of fluxes).
The additional pivot magnitude used in the experiment serves to remove the degeneracy in the luminosity class for a
specific galaxy type.

Individual experiments as well as their comparison with results provided by others, were evaluated in a consistent
and objective manner using a homogeneous and standard set of statistical indicators:
\begin{itemize}
\item the bias, defined as the mean value of the residuals $\Delta z = z_{spec} - z_{phot}$;
\item  the standard deviation ($\sigma$) of the residuals;
\item the normalized median absolute deviation or $NMAD$ of the residuals, defined as
$NMAD(\Delta z) = 1.48 \times Median\left( \left| \Delta z \right| \right )$;
\item all the above quantities calculated also on the normalized residuals, i.e.
$\Delta z_{norm} = \frac{z_{spec} - z_{phot}}{1+z_{spec}}$.
\end{itemize}
Furthermore, as an overall estimate of the accuracy of the final results it can be used the prescription in \citealt{clsi2006} by deriving
the {\it overall uncertainty} (OU) defined as $bias_{norm} \pm \sigma_{norm}$.

Results are given in Table~\ref{tabres}, where we also compare our results with those obtained
by \cite{laurino11}, who used on the SDSS Data Release 7 objects, a machine learning model with a slightly more complex
architecture, named WGE (Weak gated Experts) and which, to the best of our knowledge, has achieved the higher accuracy so far.

\begin{table*}
\tiny
\begin{center}
\begin{tabular}{lccccccccc}
\hline
ref             & $|bias|$ &  $\sigma$ &  NMAD &  rms  &$|bias_{norm}|$& $\sigma_{norm}$  &NMAD$_{norm}$  &RMS$_{norm}$ \\
\hline
Overall               		& 0.000030 & 0.028 & 0.0196  & 0.028 & 0.000574 & 0.023 & 0.0172 & 0.023\\
$\left[0,z_1\right]$  		& 0.008259 & 0.022 & 0.0179  & 0.023 & 0.007895 & 0.020 & 0.0167 & 0.022\\
$\left]z_1,z_2\right]$		& 0.008975 & 0.023 & 0.0195  & 0.025 & 0.007828 & 0.020 & 0.0171 & 0.022\\
$\left]z_2,z_3\right]$		& 0.006664 & 0.035 & 0.0234  & 0.035 & 0.005640 & 0.028 & 0.0188 & 0.029\\
$ > z_3$              		& 0.008568 & 0.037 & 0.0255  & 0.038 & 0.005705 & 0.024 & 0.0177 & 0.025\\
$\left[0.05,0.6\right]$   & 0.002224 & 0.026 & 0.0188  & 0.026 & 0.001643 & 0.022 & 0.0163 & 0.022\\
Laurino et al.       	 	  & 0.015    & 0.015 & 0.016   & 0.021 & 0.014    & 0.013 & 0.013  & 0.019\\
\hline
& $|\Delta z|$ & $|\Delta z|$ & $|\Delta z|$ & $|\Delta z_{norm}|$ &$ |\Delta z_{norm}|$ & $ |\Delta z_{norm}|$ & skewness &kurtosis\\
& $> 0.15$     & $> 1 \sigma$ & $> 2 \sigma$ & $> 0.15$            & $> 1 \sigma$        & $> 2 \sigma$\\
\hline
Overall               		& 0.12\% &22.61\% & 5.16\%& 0.04\%& 23.57\%& 5.43\%& $ 1.37\times 10^{-16}$& 1.8\\
$\left[0,z_1\right]$  		& 0.03\% &27.30\% & 6.56\%& 0.02\%& 27.02\%& 6.62\%& $ 9.73\times 10^{-17}$& 1.8\\
$\left]z_1,z_2\right]$		& 0.02\% &28.13\% & 7.16\%& 0.01\%& 28.28\%& 7.20\%& $ 1.23\times 10^{-15}$& 1.8\\
$\left]z_2,z_3\right]$		& 0.36\% &22.51\% & 6.57\%& 0.08\%& 22.54\%& 6.67\%& $ 8.03\times 10^{-16}$& 1.8\\
$ > z_3$              		& 0.50\% &20.56\% & 4.17\%& 0.16\%& 22.02\%& 4.34\%& $-1.13\times 10^{-15}$& 1.8\\
$\left[0.05,0.6\right]$     	& 0.11\% &22.84\% & 5.44\%& 0.03\%& 23.62\%& 5.51\%& $ 2.24\times 10^{-15}$& 1.8\\
\hline
\end{tabular}
\end{center}
\caption{Results for the \textit{MIXED} experiment (see text). The table is split in two parts. In the upper half we provide standard statistical indicators (see text for an explanation) used to evaluate the performances of photo-z methods. We also include the same indicators for \cite{laurino11}. In the lower half of the table we report the fraction (in percentage) of outliers computed using a fixed threshold of $0.15$, the more meaningful $1$, and $2$ $\sigma$ clipping thresholds and the values of skewness and kurtosis of the $\sigma \left( \Delta z_{norm} \right)$ distributions. The redshift range starting from $z=0$ up to the redshift $z_1=0.115$ includes $50\%$ of the objects in the test set; the second one using the range from $z_1=0.115$ to $z_2=0.177$ corresponds to an additional 25\% of the sample; the third one using the range $z_2=0.177$ to $z_3=0.345$ corresponds to an additional 15\% and, finally, the fourth one includes all remaining objects (redshift $> z_3 = 0.345$). The last row reports the statistics in the redshift range $\left[0.05,0.6\right]$ corresponding to the same range covered in \cite{laurino11}} \label{tabres}
\end{table*}

In the second half of the table we give the fraction of outliers, i.e. the fraction of objects for which the photometric redshift estimate deviates more than $0.15$ in absolute value, or deviating more  than $1$, or $2\sigma$ from the spectroscopic value.

Summarizing, MLPQNA achieves the very small bias of $\sim 3 \times 10^{-5}$, and a normalized standard deviation of $0.023$.  Moreover, our method leads to a very small fraction of outliers, i.e. less than 0.04\% and $\sim 5.4\%$ using the $\left| \Delta z_{norm} \right| > 0.15$ and the $2\sigma$ criteria, respectively.

\begin{figure*}
   \centering
   \includegraphics[width=14cm]{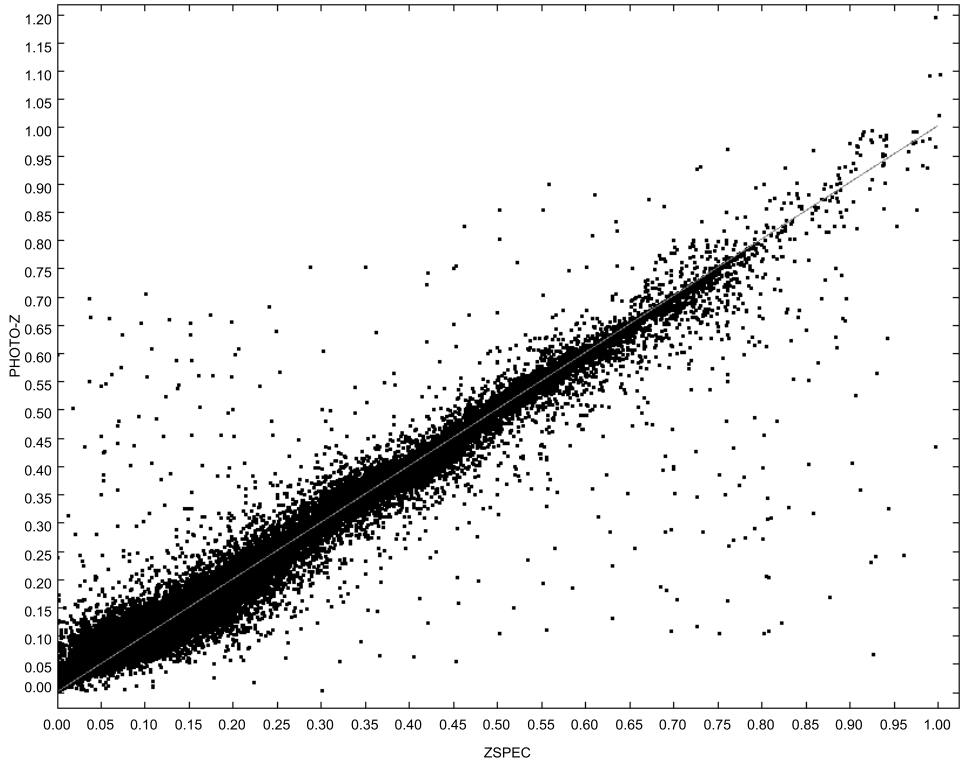}

	\caption{Spectroscopic versus photometric redshifts for SDSS $DR9$ galaxies of the blind test set.}\label{scattertest}
    \end{figure*}

In Fig.~\ref{scattertest}, we plot the photometric redshift estimates versus the spectroscopic redshift values for all objects in the test set.
After the rejection of catastrophic outliers, as defined by the $\left| \Delta z_{norm} \right| > 2 \sigma \left( \Delta z_{norm} \right)$, we obtain a $\sigma_{norm}$ of $\sim 0.0174$, which is larger than $NMAD_{norm}$.
This is exactly what has to be expected according to \cite{mobasher2007}. In fact, in the case where photo-z are empirical, it is always useful to analyze the direct correlation between the $NMAD_{norm}$ and the standard deviation $\sigma_{norm}$ calculated on data which are not catastrophic outliers. In these cases, a correct photo-z prediction occurs whenever the quantity $NMAD_{norm}$ is lower than the $\sigma_{norm}$ for the \textit{cleaned} sample.

It needs to be noticed that for empirical methods the $\sigma \left( \Delta z_{norm} \right)$ overestimates the theoretical gaussian $\sigma$, mainly due to catastrophic outliers and to the intrinsic training error.

In order to better characterize the performances of the experiment, we computed also the statistics on subsets of the test data binned accordingly to either redshift or magnitude range.

For what redshifts are concerned, as it is shown in Table.~\ref{tabres} we built the first subset by using objects in the redshift range starting from $z=0$ up to the redshift $z_1=0.11549$ which includes $50\%$ of the objects in the test set; the second one using the range from $z_1=0.115$ to $z_2=0.177$ (corresponding to an additional 25\% of the sample); the third one using the range $z_2=0.177$ to $z_3=0.345$ (corresponding to an additional 15\%) and, finally, the fourth one included all remaining objects (redshift $> z_3 = 0.345$). We also derived statistics in the redshift range $\left[0.05,0.6\right]$ corresponding to the same range covered in \cite{laurino11} (in order to allow a fair comparison).

The behaviour of the residuals as a function of the magnitude in the SDSS $r$ band was instead studied in the three bins listed in Table~\ref{TAB:magerr}.
\begin{table*}\label{TAB:magerr}
\begin{center}
\begin{tabular}{lccccccc}
\hline
mag bin (r)                & test objects & $|bias_{norm}|$  & $\sigma_{norm}$ & $quality flag$ & skewness &kurtosis\\
\hline
All			               & 347,342 		& 0.000574     & 0.023 		   & -	& $ 1.37\times 10^{-16}$& 1.8\\
$\left[ 15.4, 20.4\right]$ & 321,514      	& 0.000539     & 0.022         & 1 	& $-1.92\times 10^{-17}$& 1.8\\
$\left] 20.4, 22.2\right]$ &  25,375      	& 0.001067     & 0.032         & 2	& $ 6.34\times 10^{-18}$& 1.8\\
$\left] 22.2, 23.0\right[$ &     453      	& 0.001745     & 0.028         & 3	& $ 4.37\times 10^{-16}$& 1.8\\
\hline
\end{tabular}
\end{center}
\caption{Test set data distribution with $r$ mag binning. The table is ordered on the $r$ mag intervals (first column). The second column reports the number of objects within the related bin. Columns 3 gives the assigned photo-z quality flag. Finally, columns 4 and 5 give, respectively, the skewness and kurtosis of the distribution of the residuals.}
\end{table*}

Using these information, we assigned a photo-z quality flag (from $1$ as \textit{best quality} to $3$ as \textit{worst quality}) to all objects in all $r$ mag bins, by following both the $r$ mag completeness limit and the $\sigma_{norm}$ trend as criteria. Results are summarized in Table~\ref{TAB:magerr}.

As expected, the $\sigma_{norm}$ error remains still acceptable also slightly outside the $r$ magnitude completeness limit ($r<22.2$). In this region, however, the number of training points is rather small and for the reasons stated above, the predicted redshifts need to be taken with some cautions since, given the selection criteria applied to select the targets for the spectroscopic survey, it is very likely that not all galaxy types are present in the knowledge base and that the much wider population of objects with photometric observations only is not well represented in the training set.

It needs however to be stressed that in the already mentioned PHAT1 contest, MLPQNA obtained very good results using a KB of size (i.e. $\sim 500$ objects) similar to that used for training in the last magnitude bin.

In order to better characterize the distribution of the residuals in terms of gaussianity of the distributions, we fitted a Gaussian to the residuals in the
various quality bins obtaining the kurtosis and skewness listed in \ref{TAB:magerr}. The distributions of residuals appear to be quite symmetric even though slightly leptokurtic.

\section{The photometric Catalogue}\label{catalog}

To produce the final catalogue we downloaded from the SDSS DR9 data server\footnote{\url{http://skyserver.sdss3.org/CasJobs/}} all objects
falling within the declination range $\left[ -30^{\circ}, +85^{\circ}\right]$ and detected in all SDSS bands.

We underline that all empirical photo-z methods suffer from a poor capability to extrapolate outside the range of distributions
imposed by the training. In other words, outside the limits of magnitudes and zspec used in the training set, these methods do not ensure optimal performances.
Hence, in order to remain in a safe condition, we performed a selection of objects in the final catalog according the same selection done on the training and test sample limits.

Furthermore, the SDSS DR$9$ hosts objects which are spectroscopically recognized as galaxies, but whose photometric class is different. In most cases such objects are photometrically
classified as stars. From the spectral point of view indeed, there is a zspec value assigned to most of such objects, although they are lost from any photometric search based on galaxy type.
Hence, for completeness reason, we added such objects to the photo-z photometric catalogue, by retrieving them through a special SQL query (see Appendix Sec.~\ref{specialquery}).

For convenience, the whole catalogue was split in $59$ files, containing a total of $143,584,848$ objects with the estimated photo-z. Among them, the file with suffix \textit{specialObjects} includes the photo-z for special objects with a mismatch between photometric and spectroscopic class assignment.
The other $58$ files forming the final catalogue correspond to different declination ranges and each is structured in $24$ columns containing:
\begin{itemize}
\item column 1: the SDSS-DR9 object identification;
\item columns 2 and 3: right ascension and declination;
\item columns 4-8: the $u$, $g$, $r$, $i$, and $z$ PSF magnitudes;
\item columns 9-13: the $psfMag\_err$ error for all magnitudes;
\item columns 14-18: the extinction for each magnitude;
\item columns 19-22: the colors derived from $ugriz$ $psfMag$ type magnitudes;
\item column 23: the estimated photo-z;
\item column 24: quality flag of the photo-z obtained from the information gathered during the analysis of the test set. $1$ stands for the best photo-z accuracy, $2$ for photo-z with lower accuracy, and $3$ for the photo-z related to objects outside the $psfMag\_r$ completeness limit.
\end{itemize}

The produced photometric catalogue is publicly available for download at the public URL:
\url{http://dame.dsf.unina.it/catalog/DR9PHOTOZ/}.

\begin{figure*}
\centering
   \includegraphics[width=15cm]{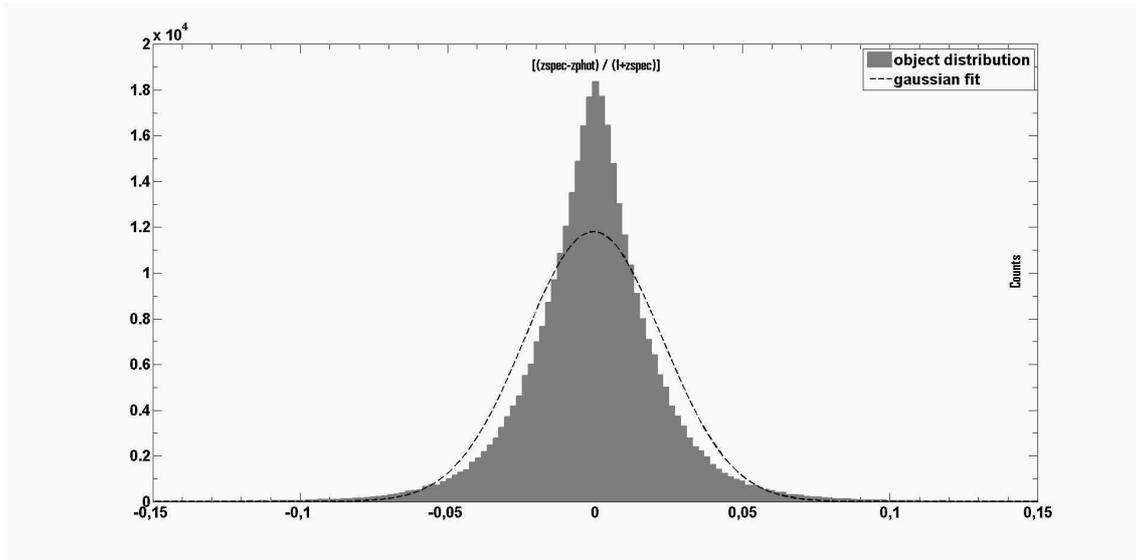}\\(a)\\
   \includegraphics[width=15cm]{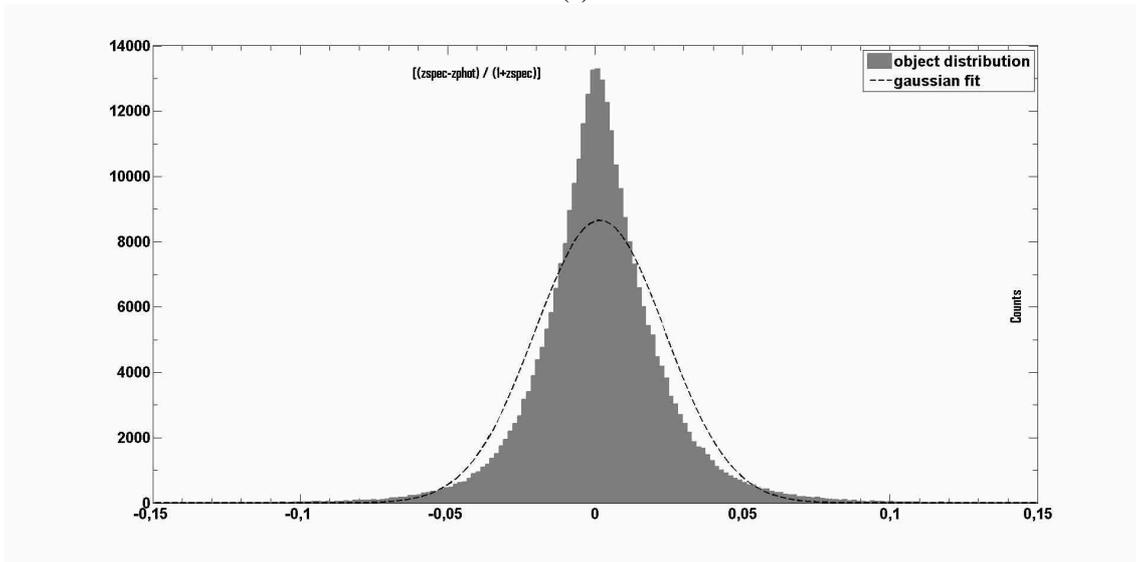}\\(b)\\

	\caption{Distribution of the residuals in two redshift bins.
	All objects (a), zspec within the range $\left[0.05, 0.6\right]$ (b).  The plotted range is $-0.15 \leq \left| \Delta z_{norm} \right| \leq 0.15$.}\label{errorsAll}
    \end{figure*}

\begin{figure*}
   \centering
   \includegraphics[width=15cm]{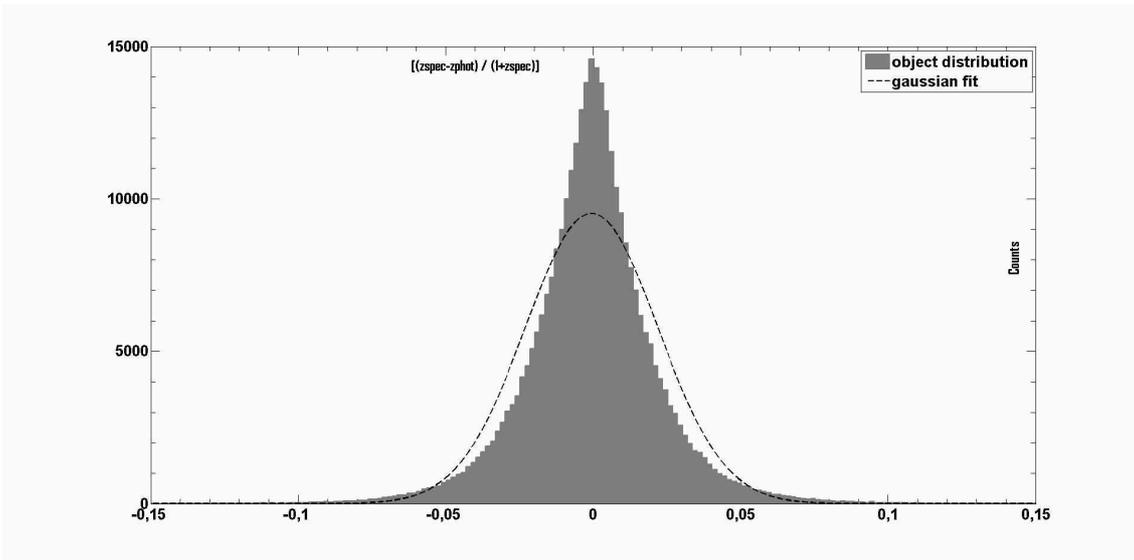}\\(a)\\
   \includegraphics[width=15cm]{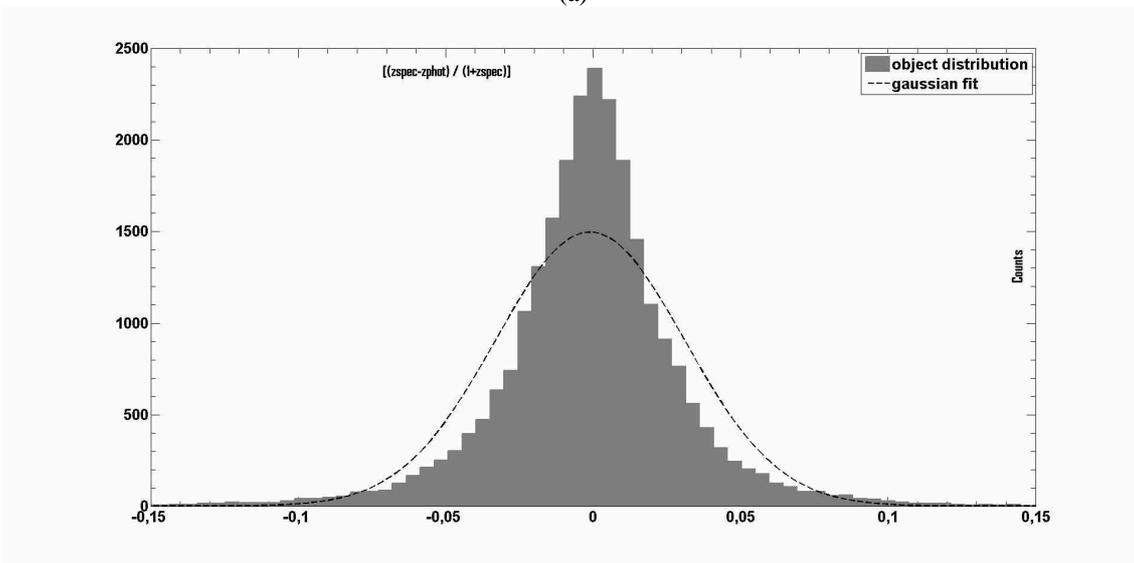}\\(b)\\
   \includegraphics[width=15cm]{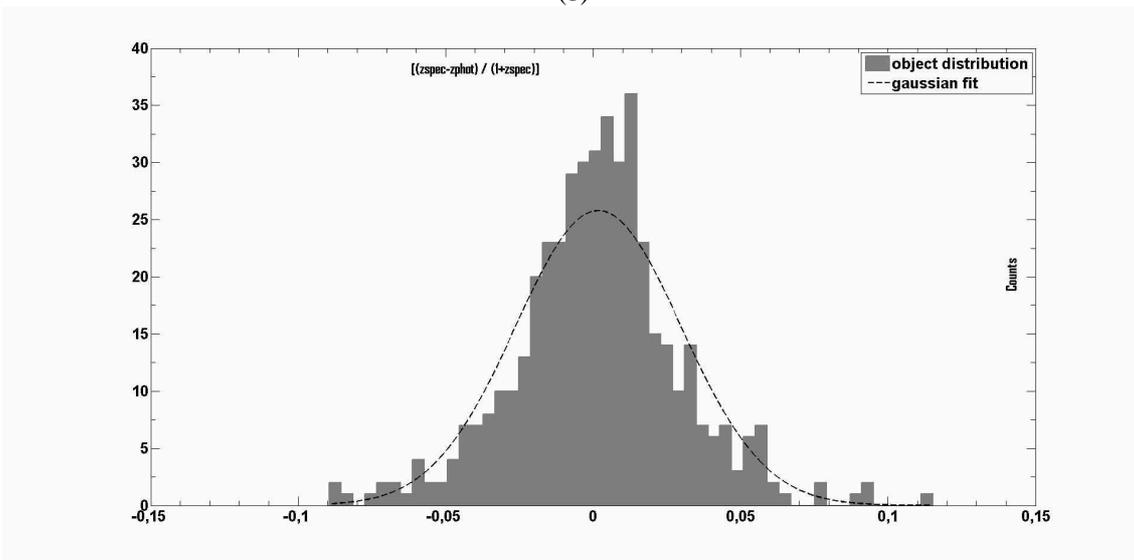}\\(c)\\
   \caption{Distribution of the residuals in three $psfMag\_r$ magnitude bins. Respectively, $\left[ 15.4, 20.4\right]$ (a), $\left] 20.4, 22.2\right]$ (b) and $\left] 22.2, 23.0\right[$  (c).  The plotted range is $-0.15 \leq \left| \Delta z_{norm} \right| \leq 0.15$.}\label{errorsMagBin}
    \end{figure*}

    \begin{figure*}
   \centering
   \includegraphics[width=15cm]{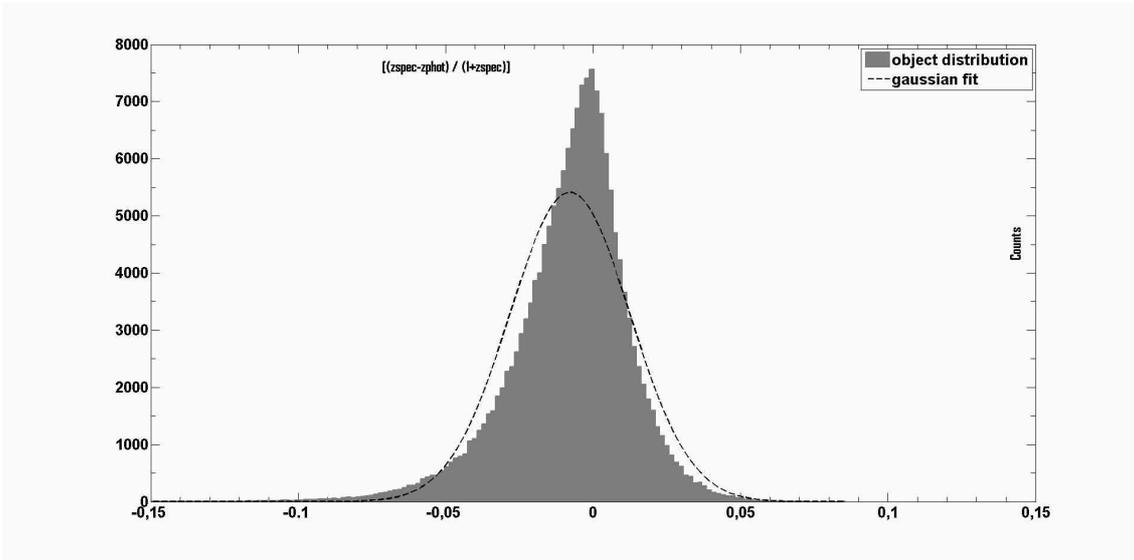}\\(a)\\
   \includegraphics[width=15cm]{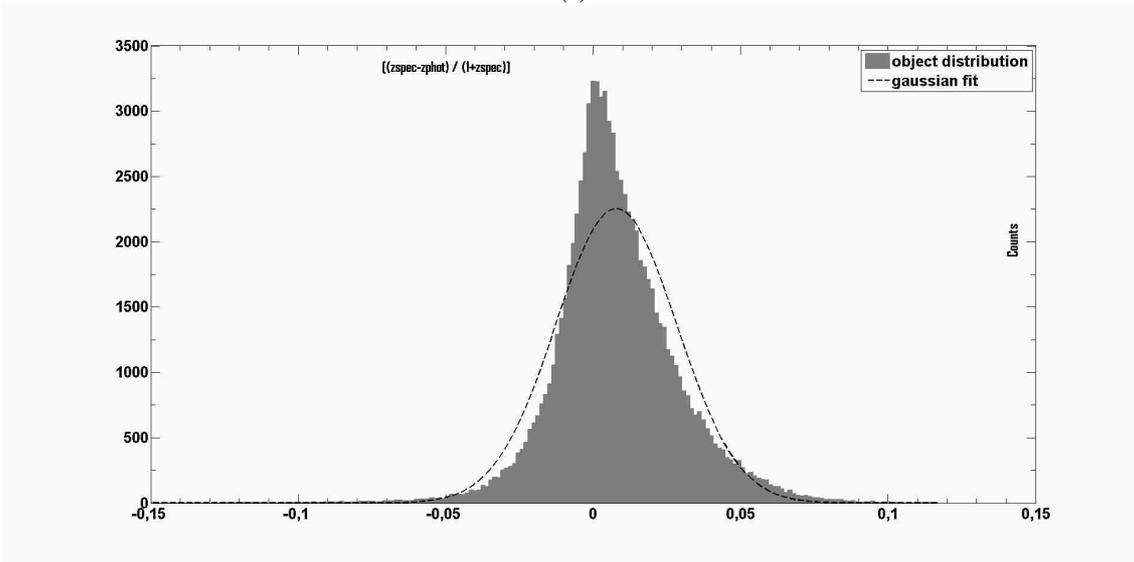}\\(b)\\

   \caption{Distribution of the residuals in zspec bins. Respectively,  $\left[0, z_1\right]$ (a), $\left]z_1,z_2\right]$ (b).  The plotted range is $-0.15 \leq \left| \Delta z\right| \leq 0.15$. The limits are, respectively, $z_1 = 0.115, z_2 = 0.177$}\label{errorsRedBin1}
    \end{figure*}

    \begin{figure*}
   \centering
   \includegraphics[width=15cm]{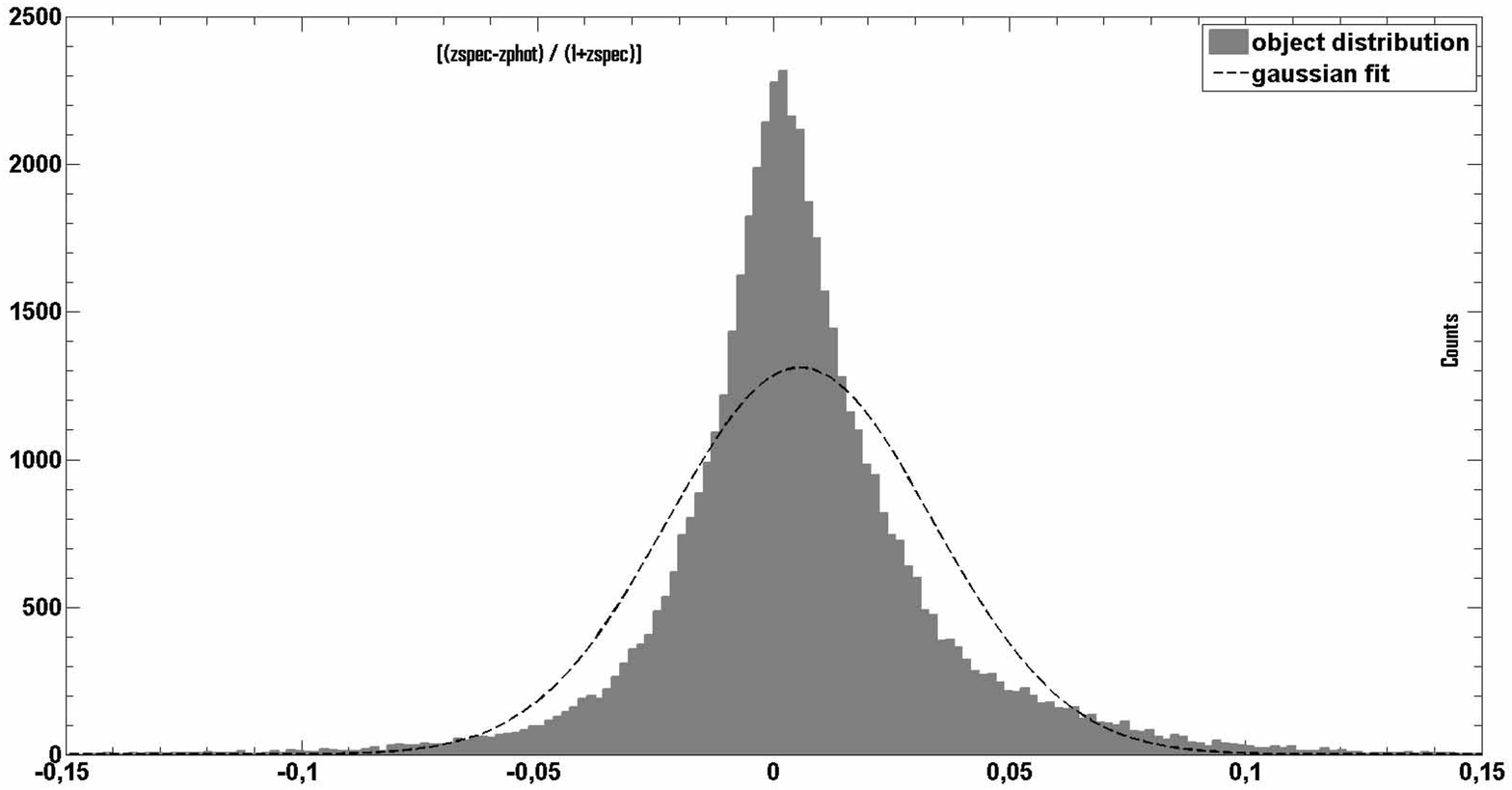}\\(a)\\
   \includegraphics[width=15cm]{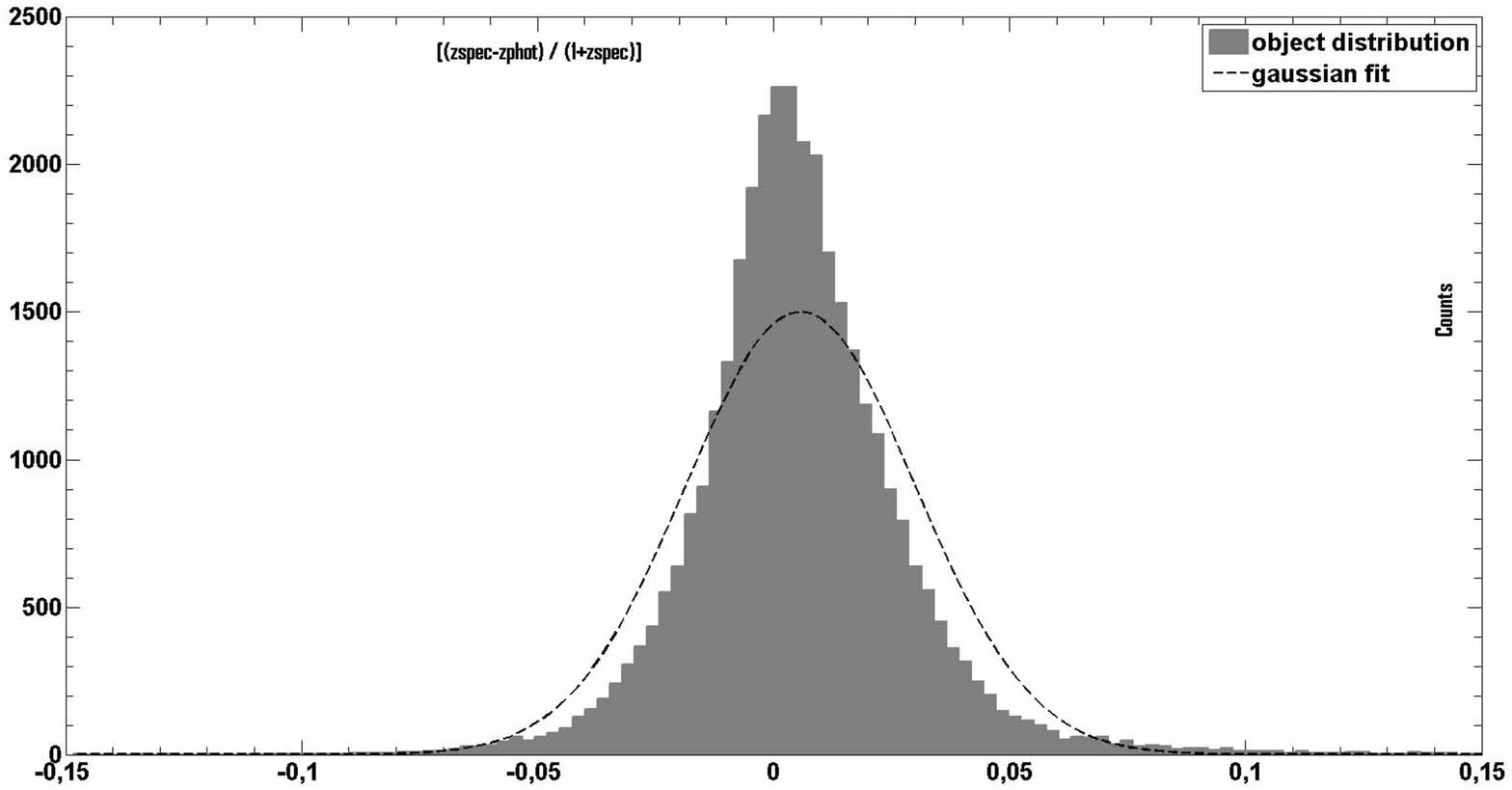}\\(b)\\

   \caption{Distribution of the residuals in zspec bins. Respectively,  $\left]z_2,z_3\right]$ (a), $ > z_3$ (b).  The plotted range is $-0.15 \leq \left| \Delta z\right| \leq 0.15$. The limits are, respectively, $z_2 = 0.177, z_3 = 0.345$}\label{errorsRedBin2}
    \end{figure*}

\section{Conclusions}\label{conclusions}

The MLPQNA neural network was applied to the SDSS-DR9 photometric galaxy data, using a knowledge base derived from the SDSS-DR9 spectroscopic subsample.

After a set of experiments the best results were obtained with a two hidden layer network, using a combination of the $4$ SDSS colors (obtained from the SDSS $psfMag$) plus the pivot magnitude $psfMag$ in the $r$ band. This gives a normalized overall uncertainty of $\sigma=0.023$ with a very small average bias of $\sim 3\times 10^{-5}$, a low $NMAD$, and to a low fraction of outliers ($\sim5\%$ at $2\sigma$ and $\sim0.1\%$ at $0.15$). After the rejection of catastrophic outliers, the residual uncertainty is $\sigma=0.0174$.

The trained network was then used to process the galaxies in the SDSS-DR9 matching the above outlined selection criteria and to produce the complete photometric catalogue. This catalog consists of photo-z estimates for more than $143$ million SDSS-DR9 galaxies and is available at the URL: \url{http://dame.dsf.unina.it/catalog/DR9PHOTOZ/}.

\begin{acknowledgements}
\noindent The authors would like to thank the anonymous referee for extremely precious comments and suggestions.
\noindent Part of this work was supported by the PRIN-MIUR 2011, \textit{Cosmology with the Euclid space mission}, and by the Project F.A.R.O., $3^{rd}$ call by the University Federico II of Naples.
\noindent One of us (GL) wishes to thank G.S. Djorgovski and the Department of Astronomy and Astrophysics at the Caltech for support and hospitality.\\
\end{acknowledgements}


\appendix
\section{Spectroscopic Query}
\label{spectroquery}

The following SQL code has been used to obtain the spectroscopic Knowledge Base to train and test the model.

\begin{verbatim}
SELECT
    p.objid, s.specObjID, p.ra, p.dec,
    p.psfMag_u, p.psfMag_g, p.psfMag_r, p.psfMag_i,
    p.psfMag_z, p.psfmagerr_u, p.psfmagerr_g,
    p.psfmagerr_r, p.psfmagerr_i, p.psfmagerr_z,
    p.fiberMag_u, p.fiberMag_g, p.fiberMag_r,
    p.fiberMag_i, p.fiberMag_z, p.fibermagerr_u,
    p.fibermagerr_g, p.fibermagerr_r,
    p.fibermagerr_i, p.fibermagerr_z,
    p.petroMag_u, p.petroMag_g, p.petroMag_r,
    p.petroMag_i, p.petroMag_z, p.petromagerr_u,
    p.petromagerr_g, p.petromagerr_r,
    p.petromagerr_i, p.petromagerr_z,
    p.modelMag_u, p.modelMag_g, p.modelMag_r,
    p.modelMag_i, p.modelMag_z,
    p.modelmagerr_u, p.modelmagerr_g,
    p.modelmagerr_r, p.modelmagerr_i,
    p.modelmagerr_z,
    p.extinction_u, p.extinction_g,
    p.extinction_r, p.extinction_i,
    p.extinction_z, s.z as zspec,
    s.zErr as zspec_err, s.zWarning,
    s.class, s.subclass, s.primTarget
INTO
    mydb.galaxies_spec
FROM
    PhotoObjAll as p,
    SpecObj as s
WHERE
    s.class = 'GALAXY' AND s.zWarning = 0 AND
    p.mode = 1 AND p.SpecObjID = s.SpecObjID AND
    dbo.fPhotoFlags('PEAKCENTER') != 0 AND
    dbo.fPhotoFlags('NOTCHECKED') != 0 AND
    dbo.fPhotoFlags('DEBLEND_NOPEAK') != 0 AND
    dbo.fPhotoFlags('PSF_FLUX_INTERP') != 0 AND
    dbo.fPhotoFlags('BAD_COUNTS_ERROR') != 0 AND
    dbo.fPhotoFlags('INTERP_CENTER') != 0
\end{verbatim}

\section{Photometric Query}
\label{photoquery}

The produced photometric catalogue with the estimated photo-z has been taken from SDSS DR9 service, by applying the following SQL query.
The reported code here is referred to a DEC range between $60$ and $65$ deg, as an example.

\begin{verbatim}
SELECT
    p.objid, p.ra, p.dec,
    p.psfMag_u, p.psfMag_g, p.psfMag_r,
    p.psfMag_i, p.psfMag_z,
    p.psfmagerr_u,  p.psfmagerr_g,
    p.psfmagerr_r,  p.psfmagerr_i,
    p.psfmagerr_z,
    p.extinction_u, p.extinction_g,
    p.extinction_r, p.extinction_i,
    p.extinction_z
INTO
    mydb.p60_p65
FROM
    Galaxy as p
WHERE
    p.mode = 1 AND
    dbo.fPhotoFlags('PEAKCENTER') != 0 AND
    dbo.fPhotoFlags('NOTCHECKED') != 0 AND
    dbo.fPhotoFlags('DEBLEND_NOPEAK') != 0 AND
    dbo.fPhotoFlags('PSF_FLUX_INTERP') != 0 AND
    dbo.fPhotoFlags('BAD_COUNTS_ERROR') != 0 AND
    dbo.fPhotoFlags('INTERP_CENTER') != 0 AND
    p.dec >= 60 AND p.dec <65
\end{verbatim}

\section{Special Query}
\label{specialquery}

Below it is reported the SQL code used for the query needed to integrate the photo-z catalogue with objects spectroscopically recognized as galaxies but photometrically assigned to different classes within the SDSS DR$9$.

\begin{verbatim}
SELECT
    p.objid, p.ra, p.dec,
    p.psfMag_u, p.psfMag_g, p.psfMag_r,
    p.psfMag_i, p.psfMag_z,
    p.psfmagerr_u, p.psfmagerr_g,
    p.psfmagerr_r,  p.psfmagerr_i,
    p.psfmagerr_z, p.extinction_u,
    p.extinction_g, p.extinction_r,
    p.extinction_i, p.extinction_z
INTO
    mydb.photoerror
FROM
    PhotoObjAll as p,
    SpecObj as s
WHERE
    s.class = 'GALAXY' AND p.type != 3 AND p.mode = 1 AND
    dbo.fPhotoFlags('PEAKCENTER') != 0 AND
    dbo.fPhotoFlags('NOTCHECKED') != 0 AND
    dbo.fPhotoFlags('DEBLEND_NOPEAK') != 0 AND
    dbo.fPhotoFlags('PSF_FLUX_INTERP') != 0 AND
    dbo.fPhotoFlags('BAD_COUNTS_ERROR') != 0 AND
    dbo.fPhotoFlags('INTERP_CENTER') != 0 AND
    p.SpecObjID = s.SpecObjID
\end{verbatim}

\end{document}